\documentclass[journal,hideappendix]{vgtc}        

\onlineid{0}

\vgtccategory{Research}

\vgtcpapertype{please specify}

\title{Shadowless Projection Mapping for Tabletop Workspaces \\ with Synthetic Aperture Projector}

\author{%
  \authororcid{Takahiro Okamoto*}{0009-0001-6709-967X},
  \authororcid{Masaki Takeuchi*}{0000-0002-0773-7705},
  \authororcid{Masataka Sawayama}{0000-0003-0725-3214}, and 
  \authororcid{Daisuke Iwai}{0000-0002-3493-5635}
}

\authorfooter{
  \item{*: Takahiro Okamoto and Masaki Takeuchi are co-first authors.}
  \item
  	Takahiro Okamoto, Masaki Takeuchi, and Daisuke Iwai are with Graduate School of Engineering Science, The University of Osaka.
  	E-mail: see https://www.xr.sys.es.osaka-u.ac.jp/.
  \item
  	Masataka Sawayama is with Graduate School of Information Science and Technology, Hokkaido University and Prometech CG Research.
  	E-mail: masataka\_sawayama@ist.hokudai.ac.jp.

}

\abstract{
Projection mapping (PM) enables augmented reality (AR) experiences without requiring users to wear head-mounted displays and supports multi-user interaction.
It is regarded as a promising technology for a variety of applications in which users interact with content superimposed onto augmented objects in tabletop workspaces, including remote collaboration, healthcare, industrial design, urban planning, artwork creation, and office work.
However, conventional PM systems often suffer from projection shadows when users occlude the light path.
Prior approaches employing multiple distributed projectors can compensate for occlusion, but suffer from latency due to computational processing, degrading the user experience.
In this research, we introduce a synthetic-aperture PM system that uses a significantly larger number of projectors, arranged densely in the environment, to achieve delay-free, shadowless projection for tabletop workspaces without requiring computational compensation.
To address spatial resolution degradation caused by subpixel misalignment among overlaid projections, we develop and validate an offline blur compensation method whose computation time remains independent of the number of projectors.
Furthermore, we demonstrate that our shadowless PM plays a critical role in achieving a fundamental goal of PM: altering material properties without evoking {\it projection-like} impression.
Specifically, we define this perceptual impression as ``sense of projection (SoP)'' and establish a PM design framework to minimize the SoP based on user studies.
} 


\teaser{
  \centering
  \includegraphics[width=\linewidth]{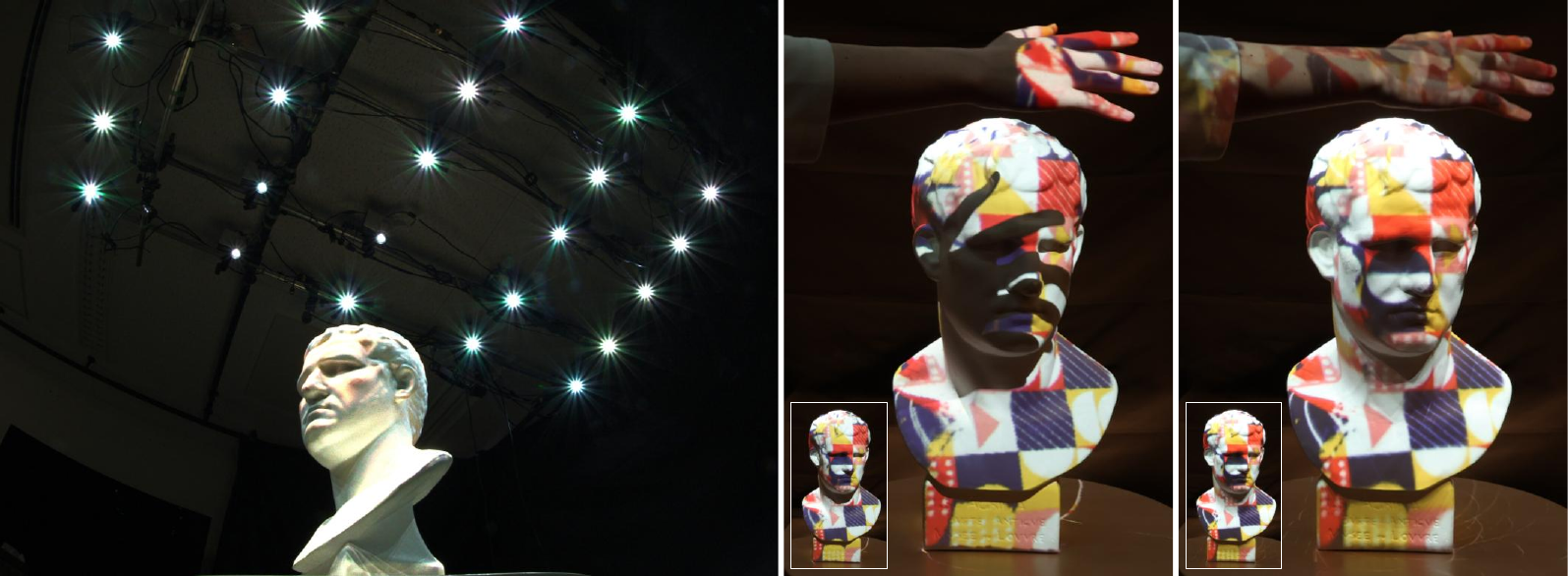}
  \caption{Shadowless projection mapping system using a synthetic aperture projector. (Left) The system consists of 25 projectors arranged in a 5×5 array mounted on the ceiling and a plaster bust placed directly below. (Center and Right) Comparison of shadow removal performance between the single-projector setup (center) and the 25-projector setup (right), both with a hand positioned above the bust at the same location. The insets show the projected appearance without the occluding hand.}
  \label{fig:teaser}
}

\graphicspath{{figs/}{figures/}{pictures/}{images/}{./}} 

\usepackage{tabu}                      
\usepackage{booktabs}                  
\usepackage{lipsum}                    
\usepackage{mwe}                       

\usepackage{mathptmx}                  
\usepackage{amsfonts}

\usepackage{comment}

\begin{document}


\firstsection{Introduction}

\maketitle

Projection mapping (PM) overlays virtual imagery onto physical surfaces to enable augmented reality (AR) experiences~\cite{10.5555/1088894,DaisukeIWAI2024pjab.100.012}.
Thanks to its advantages---such as not requiring users to wear head-mounted displays (HMDs) and supporting multi-user interaction---PM has been explored in a variety of applications, including remote collaboration~\cite{10.1145/2207676.2207704,8172039}, healthcare~\cite{00000658-201806000-00024}, industrial design~\cite{8797923,https://doi.org/10.1111/j.1467-8659.2011.02066.x,6949562,CASCINI2020103308,10937893}, urban planning~\cite{10.1145/302979.303114}, artwork creation~\cite{10.1145/2366145.2366176,10.1145/1166253.1166290,970539}, and office work~\cite{10.1145/1180495.1180519,10.1145/504704.504706}.
In many of these applications, users typically interact with content superimposed onto an augmented object within a tabletop workspace, observing the projection from various distances and angles around the object.
However, this configuration faces a technical challenge: projected light is often occluded by the user's body, resulting in shadows and incomplete projection outcomes.
To address this issue, previous studies have proposed systems that spatially distribute multiple projectors throughout the environment~\cite{https://doi.org/10.1111/cgf.13541,7164338,https://doi.org/10.1111/cgf.13085,Nagase2011,Iwai2014,990943,964509,1272728,9714121}.
When the projection from one projector is blocked and a shadow appears on the object, another, unoccluded projector compensates by projecting onto the same area.
However, this approach suffers from noticeable compensation failures when occluders move, due to system latency introduced by the compensation computation.
These failures degrade the overall AR experience.

To solve the latency problem in PM shadow removal, researchers have explored optical approaches.
Hiratani et al. demonstrated the effectiveness of an optical method by projecting imagery from multiple directions onto each point of the target surface using a large-aperture optical element (approximately 500$\times$500 mm)~\cite{10049693}.
This setup achieved delay-free, shadowless PM; however, image quality was degraded by stray light, which significantly reduced contrast.
Kusuyama et al. addressed this image degradation issue by integrating a standard-aperture projector with the large-aperture system in a coaxial configuration, thereby balancing shadowless performance with improved image quality~\cite{10670583}.
Nevertheless, these optical approaches are fundamentally limited in projection area to approximately the size of a clenched fist, due to the physical constraints imposed by the aperture size of existing optical elements, making them unsuitable for broader tabletop applications.

\begin{figure}[tb]
  \centering
  \includegraphics[width=\linewidth]{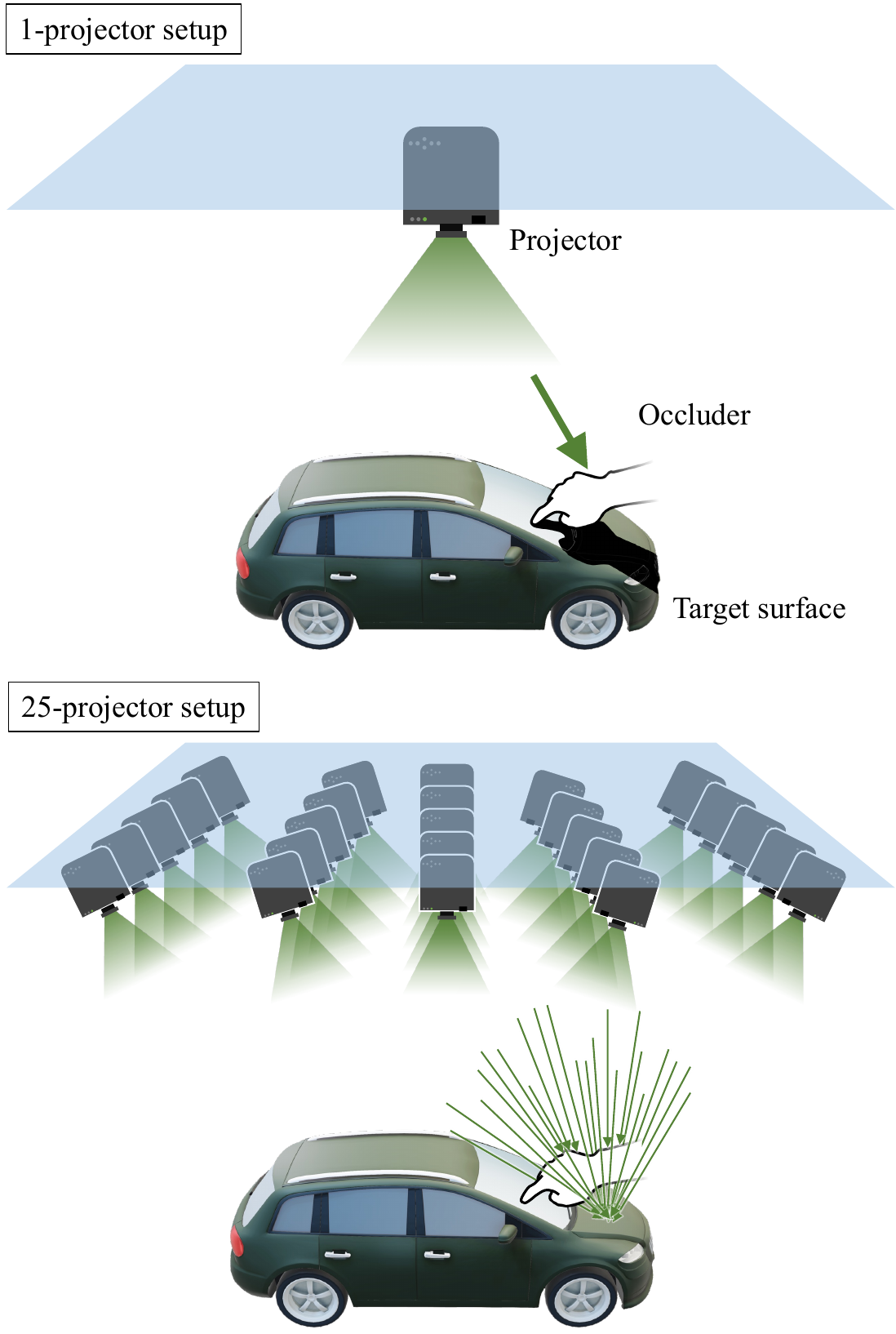}
  \caption{Proposed shadow removal principle by synthetic aperture projection. (Top) In a single-projector setup, shadows are cast when a user’s hand or other object approaches the projection surface and blocks the light path. (Bottom) In a synthetic aperture projection setup (25 projectors in this example), even if some projectors are occluded, light from the remaining projectors still reaches the surface, effectively reducing shadow formation.}
  \label{fig:concept}
\end{figure}

To overcome this limitation, we propose a shadowless PM system for tabletop workspaces using a synthetic aperture approach, effectively eliminating aperture size constraints.
Specifically, we deploy 25 projectors arranged in a 5$\times$5 grid on the ceiling, projecting the same imagery from multiple directions onto each point of the target surface (Fig.~\ref{fig:concept}).
This configuration ensures that even if some projectors are occluded by an observer, the majority remain unblocked, thereby enabling a shadowless PM experience.
When the number of projectors is sufficiently large and their directional coverage is wide enough, the entire system can be regarded as a single, large-aperture projector---capable of eliminating shadows without computational compensation.
To achieve this, precise geometric alignment of all projected images is required.
However, even if pixel-level alignment is achieved, the fact that each projector operates from different positions and angles causes projected pixels to vary in shape.
This leads to inevitable subpixel misalignments among overlapping images, resulting in undesirable blurring.
Although existing blur compensation techniques can address this degradation in spatial resolution, they require inverting a large matrix proportional to the square of the number of pixels per projector.
As a result, computation time becomes prohibitively long when using many projectors.
To address this, we propose an efficient blur compensation method whose computation time remains constant regardless of the number of projectors, and we demonstrate its accuracy through experimental validation.

In addition, during a preliminary study using our prototype, we discovered an intriguing perceptual phenomenon.
With conventional PM using a single projector, visual changes on a target surface are typically perceived as alterations in the projected image itself.
In contrast, with our synthetic aperture PM system, observers perceived the changes as alterations in the reflectance properties of the actual physical surface.
Originally, the goal of PM was to alter the perceived reflectance of real-world surfaces through projected imagery~\cite{10.1007/978-3-7091-6242-2_9}.
To achieve this, prior research has focused on geometric and photometric compensation techniques aimed at accurately reproducing target color values measured by cameras or spectroradiometers~\cite{https://doi.org/10.1111/j.1467-8659.2008.01175.x,https://doi.org/10.1111/cgf.13387}.
However, even when measured physical values match the target, human vision often interprets the result as merely a change in projected imagery rather than a change in material appearance.
This {\it projection-like} impression challenges a core assumption of PM.
To our knowledge, few studies have explicitly investigated this phenomenon---what we refer to as the ``sense of projection (SoP).''
In this study, we aim to identify system design guidelines that suppress the SoP, allowing users to perceive changes in reflectance properties rather than simply recognizing the presence of projected imagery through user studies.

Our primary contributions are that we:
\begin{itemize}
    \item Introduce a delay-free PM system for tabletop workspaces using a 25-projector synthetic aperture setup, enabling shadowless projection without the need for computational compensation,
	\item Develop and validate a blur compensation method whose computation time is independent of the number of projectors, addressing spatial resolution degradation caused by subpixel misalignment in overlaid projections,
	\item Discover a novel perceptual phenomenon---SoP---in which users perceive apparent changes in PM as alterations in the projected imagery rather than changes in the material properties, and
    \item Establish a PM design framework aimed at minimizing SoP, based on insights from user studies.
\end{itemize}

\section{Related Work}

\subsection{Shadowless Projection Mapping}

Shadow removal in PM has been an ongoing research topic since the early 2000s, yet a fundamental solution has not been achieved.
The primary challenge lies in the fact that PM is ultimately perceived by human observers.
Because of this, post-processing techniques to remove shadows after projection are not acceptable; shadow removal must occur in real time, at the moment of viewing.

Most previous approaches to shadow removal have involved placing a small number of projectors at spatially distributed positions within the environment.
These systems detect occluders~\cite{https://doi.org/10.1111/cgf.13541,4270468,Iwai2014,9714121,Nagase2011} or shadow regions~\cite{990943,1272728,964509,5544603,7164338,https://doi.org/10.1111/cgf.13085,10.1145/1015706.1015806} using cameras or depth sensors and then compute which projectors remain unoccluded, projecting compensatory imagery from them.
Such techniques are effective when the occluders are static, often producing a perception similar to that of shadow-free rear-projection system~\cite{10.1145/1095034.1095054,10.1145/1449715.1449729,7223326}.
However, when occluders are in motion, these methods require a certain amount of time for the entire pipeline---from detection of the occluder or shadow region to generation and projection of the compensatory image.
This latency prevents complete shadow elimination.
The human visual system is highly sensitive to such delays; even a few milliseconds of latency can result in noticeable image degradation~\cite{10.1145/2380116.2380174,10316453}.

To address this, recent studies have explored optical solutions rather than the computational ones~\cite{10049693,10670583}.
By employing a large-aperture optical element as the projector's objective lens, it becomes possible to project imagery onto the target surface from a wide range of directions.
If the optical element is sufficiently large relative to the size of the occluding object---such as the user's hand---then projected rays are not completely blocked, and no shadow appears in the final projected result.
Since this process requires no computational compensation, it produces no perceptible delay.
However, these studies typically employed commercially available large-aperture optical elements, with the largest being approximately 500$\times$500 mm in size.
Experimental results revealed that such systems can only remove shadows for target surfaces roughly the size of a clenched fist.
Thus, delay-free shadow removal over larger projection areas, such as tabletop workspaces envisioned in many practical applications, has yet to be realized.

\subsection{Blur Compensation in Projection Mapping}

PM suffers from defocus blur due to the typically shallow depth of field of projectors.
It is also affected by other types of blur, such as inter-reflection and subsurface scattering.
To mitigate image degradation caused by these effects, techniques have been proposed that pre-enhance the high spatial frequency components of the projected image~\cite{1640992,10.1145/1179352.1141974,4270463,10.1145/2508363.2508416}.
Blur can be mathematically modeled as a convolution of the original image with a point spread function (PSF), which characterizes the blur as an impulse response.
By inverting this linear system, a compensated image with enhanced high-frequency content can be computed.

However, in PM scenarios where the target surface is often non-planar, the PSF becomes spatially non-uniform, significantly increasing computational complexity.
Specifically, existing compensation techniques represent the PSF as a matrix with dimensions equal to the square of the number of pixels and compute its inverse.
Moreover, due to constraints in the projector's dynamic range---such as the inability to emit negative light---iterative optimization methods, such as gradient descent, are required, leading to considerable computation time.
While recent studies have achieved faster computation using deep neural networks~\cite{Kageyama:20,9714047,10400947}, in multi-projector systems like the one addressed in this study, computation time still increases linearly with the number of projectors.

In near-eye displays, multi-PSF inverse problems have been accelerated using compact kernel decomposition formulations~\cite{ebner2024gaze-contingent, mercier2017fast}. These formulations are typically derived for shift-invariant optical PSFs and focus-dependent or layered display models, rather than spatially varying blur on physical surfaces. Therefore, they do not directly address pixel-misalignment-induced blur in multi-projector projection mapping, limiting their direct applicability.

\subsection{Sense of Projection}

In PM, accurately reproducing the target visual appearance for human observers requires an understanding of how the projected results are visually perceived.
For example, in readiometric compensation techniques aimed at reproducing desired colors on textured surfaces, not only the Peak Signal-to-Noise Ratio (PSNR) of RGB values captured by a camera is used, but also perceptually informed metrics such as Structural Similarity Index Measure (SSIM) and Learned Perceptual Image Patch Similarity (LPIPS)~\cite{8578166}, which better reflect human visual characteristics~\cite{deng2024gsprocamsgaussiansplattingbasedprojectorcamera}.
In dynamic PM, where imagery must be projected onto moving objects without geometric misalignment, prior work has shown that the perceptual threshold for misalignment caused by system latency is on the order of a few milliseconds~\cite{10.1145/2380116.2380174}.
Accordingly, techniques have been developed to keep projection latency below this threshold~\cite{7516689,10.1145/3680528.3687624}.
However, these findings are largely based on psychophysical experiments using images presented on conventional computer monitors or flat, rectangular surfaces.
Recent studies suggest that PM has distinct perceptual properties that differ from those of planar, rectangular displays~\cite{10458352}.

Studies that explicitly address the SoP remain limited.
Traditionally, it has been assumed that if a pattern on a surface under white light and the same pattern projected onto a white surface yield matching sensor measurements (e.g., from a camera), then PM has successfully reproduced the original appearance~\cite{10.1007/978-3-7091-6242-2_9}.
However, in the course of our preliminary investigation, we observed that human vision may interpret such a result not as a change in the surface's reflectance properties, but simply as a change in the projected image itself---even when the sensor readings match.
To date, no studies have explicitly tackled this perceptual issue, and technical guidelines for how to design PM systems that truly achieve the original goal of altering perceived reflectance remain undeveloped.

\subsection{Our Contribution}

In this study, we adopt a synthetic aperture approach and introduce a projection framework in which a densely and widely distributed array of projectors in the environment casts imagery onto a target surface.
This enables shadow removal in PM within tabletop-scale workspaces, without incurring latency.
Specifically, we achieve this by overlaying identical imagery from 25 ceiling-mounted projectors toward the target surface from multiple directions.
However, slight subpixel misalignments between the projections from different projectors inevitably occurs, resulting in the blur in the projected result.
To address this, we develop a blur compensation technique whose computation time remains constant regardless of the number of projectors, and we experimentally validate its effectiveness.
Finally, through user studies, we demonstrate that shadow removal contributes to the suppression of the newly identified perceptul phenomenon---SoP.
We also derive design guidelines regarding the number and arrangement of projectors necessary to present projection imagery that is perceived not as projected light, but as a change in the surface's reflectance properties---thereby aligning with the original goal of PM.

\section{Synthetic Aperture Projector for Shadowless Projection Mapping}\label{sec:method}

Synthetic aperture is a technique that combines multiple small-aperture imaging or display systems to virtually construct a single large-aperture system.
Applying this principle to projection display systems theoretically enables shadowless PM.
Specifically, we propose a method where the same image is projected onto the target surface from all projectors with precise geometric alignment.
However, if the number of projectors is small and they are sparsely distributed over the environment, the contribution of each projector to the intensity of the projected result is large.
This results in a significant drop in the light intensity of the projected result when a projector is occluded.
To address this issue, prior studies have attempted to compensate for brightness loss by increasing the output of the remaining unoccluded projectors.
However, this compensation process introduces latency, potentially resulting in noticeable artifacts in the projected result.
To solve this problem, we propose using a significantly larger number of projectors, arranged more densely in the environment, compared to prior systems.

When the projections from multiple projectors overlap on the target surface, the pixel shapes from different projectors do not match, and they are misaligned at the subpixel level, resulting in blurred projections.
Mathematically, this blur can be modeled as a convolution of the input image to the projector with a PSF that represents the spatial spreading of each pixel.
Assuming that correspondences between the pixels of all projectors and those of a camera (used as a proxy for the observer) have been obtained in advance using a spatial pattern (e.g., graycode pattern) projection technique, the following discussion uses the pixel coordinates $(x,y)$ of the camera, where $x = 1, 2, \ldots, w$ and $y = 1, 2, \ldots, h$.

Suppose only the $i$-th projector illuminates the target surface.
Let $p_i(x,y)$ denote the pixel value of the input image to the projector at $(x,y)$, and $c(x,y)$ denote the corresponding pixel value captured by the camera.
Because projectors in PM are typically not oriented perfectly perpendicular to the target surface, the PSF is spatially non-uniform.
To represent this spatial non-uniformity, we denote the PSF as $k_i^{(x,y)}(u,v)$.
Then, the captured projection result can be expressed as:
\begin{equation}
  c(x,y)=(p_i*k_i^{(x,y)})(x,y),
\end{equation}
where $*$ denotes the convolution operator.
This spatially varying convolution can be rewritten in matrix form by stacking the projector input pixels into a vector $\mathbf{p}_i\in\mathbb{R}^{wh}$ and the corresponding camera pixels into a vector $\mathbf{c}\in\mathbb{R}^{wh}$, leading to:
\begin{equation}\label{eq:forward_i}
  \mathbf{c}=\mathbf{L}_i\mathbf{p}_i,
\end{equation}
where
\begin{eqnarray}
  \mathbf{p}_i=[p_i(1,1), p_i(1,2), \ldots, p_i(w,h)]^\top,\\
  \mathbf{c}=[c(1,1), c(1,2), \ldots, c(w,h)]^\top,
\end{eqnarray}
and $\mathbf{L}_i \in \mathbb{R}^{wh\times wh}$ is the light transport matrix from projector $i$ to the camera.
Because multiple projectors overlap on the target surface, Equation \ref{eq:forward_i} can be generalized to incorporate the contributions from $N$ projectors as:
\begin{equation}\label{eq:forward_sum}
  \mathbf{c}=\sum_i^N\{\mathbf{L}_i\mathbf{p}_i\}.
\end{equation}

Our goal is to find the input image $\mathbf{p}_i'$ that compensates for the blur to produce a desired appearance $\mathbf{c}'$.
A na\"{i}ve but proven-to-be-accurate solution distributes the desired appearance evenly across all projectors and computes the input image for each projector to reproduce it.
Specifically:
\begin{equation}\label{eq:inverse_naive}
  \mathbf{p}_i'=\mathbf{L}_i^{-1}\frac{\mathbf{c}'}{N}.
\end{equation}
%
In practice, due to the dynamic range limitations of projectors---specifically, the inability to output negative values and the presence of a maximum brightness limit---we solve the inverse problem using a constrained steepest descent  method based on the light transport matrix ~\cite{10.1145/1141911.1141974}.
Note that each projector accepts discrete inputs, and quantization can affect the reproduced luminance.
In our implementation, we use measured per-projector luminance response curves to map the intended luminance to projector-specific inputs. Thus, increasing the number of projectors does not reduce luminance gradation, and the main trade-off arises only when the required inputs hit the dynamic-range limits.
This computational cost is already substantial; more critically, it increases linearly with the number of projectors.

On the other hand, because the same image is projected from all projectors, we can rewrite Equation \ref{eq:forward_sum} as:
\begin{equation}
  \mathbf{c}=\mathbf{L}\mathbf{p},
\end{equation}
where $\mathbf{L}$ and $\mathbf{p}$ represent the sum of all projectors' light transport matrices (i.e., $\mathbf{L} = \sum_i^N{\mathbf{L}_i}$) and the input image to all the projectors, respectively.
Then, the input image compensating the blur to produce a desired appearance can be computed as:
\begin{equation}\label{eq:inverse_proposed}
	\mathbf{p}'=\mathbf{L}^{-1}\mathbf{c}'.
\end{equation}
Therefore, the computational cost remains constant regardless of the number of projectors in our method.

\section{Experiments}

We built a two-dimensional projector array to implement synthetic aperture projection and evaluated its shadow removal performance and the effectiveness of the proposed blur compensation method.

\subsection{System Configuration}\label{subsec:system}

\begin{figure}[!t]
    \centering
    \subfloat[]{\includegraphics[width=\linewidth]{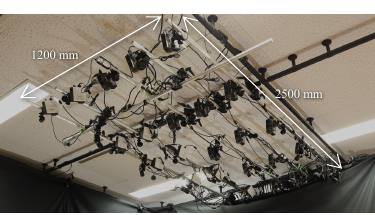}%
    \label{fig:system_overview_bright}}
    \hfil
    \subfloat[]{\includegraphics[width=\linewidth]{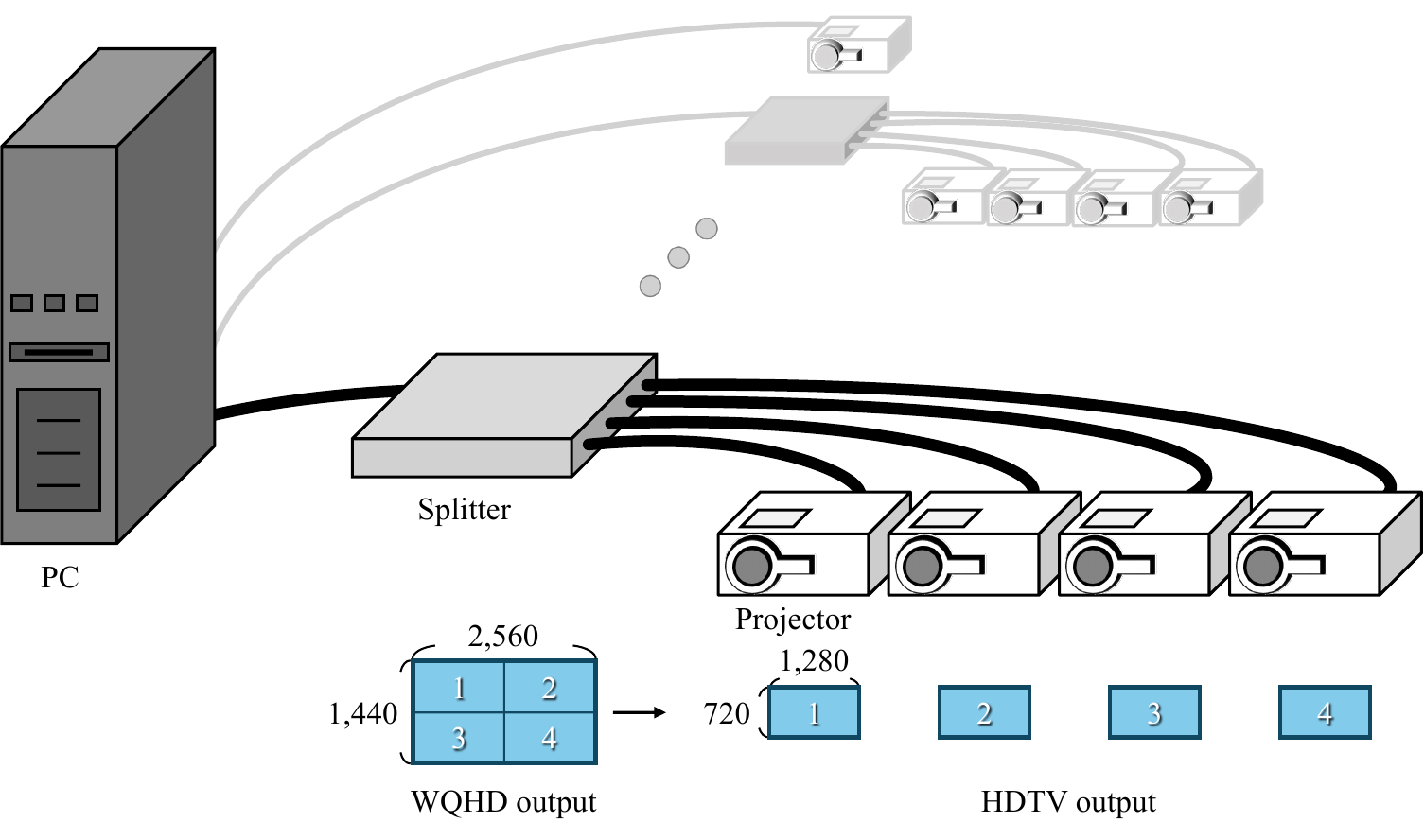}%
    \label{fig:system_illumi}}
    \caption{The prototype synthetic aperture projection system using 25 projectors arranged in a 5$\times$5 grid. (a) A captured photograph of the ceiling-mounted projectors. (b) The system architecture that enables 2K video output to the 25 projectors.}
    \label{fig:system_setup}
\end{figure}

We installed 25 projectors (Optoma, ML1050ST+S1J) in a 5$\times$5 two-dimensional array on the ceiling (Fig.~\ref{fig:system_overview_bright}).
The number 25 was determined based on practical constraints.
On a typical PC, the number of available PCI Express $\times$16 slots is limited.
On the PC we used, only two graphics cards could be installed.
Each of our graphics cards (NVIDIA, GeForce RTX 3070) had four DisplayPort ports capable of 4K output.
By using a splitter (AJA, HA5-4K), the 4K image output from each port could be divided into four 2K outputs, enabling control of up to 32 projectors at 2K resolution (Fig.~\ref{fig:system_illumi}).
Considering later user studies in Section \ref{sec:user-study} where we vary the number and density of projectors, we chose the number of projectors to be a square of an integer, and selected a 5$\times$5 configuration.

We aligned the projected images from the projectors on the target surface as follows.
Different calibration procedures were applied depending on whether the target surface was planar or non-planar.
For planar surfaces, we used a single ceiling-mounted camera (Basler, a2A2590-60ucBAS) to capture the correspondence between the world 2D coordinates $(x_w, y_w)$ defined on the target plane and camera image coordinates $(x_c, y_c)$, and calibrated the homography matrix $\mathbf{H}_{w2c}$ between them.
Each projector was calibrated with the camera using Gray-code projection to obtain the homography matrix $\mathbf{H}_{c2p_i}$ mapping camera image coordinates to each projector image coordinates $(x_{p_i}, y_{p_i})$.
Using the multiplied homography $\mathbf{H}_{w2c}\mathbf{H}_{c2p_i}$, we aligned each projector's output onto the desired rectangular area on the surface.

For non-planar surfaces, we first scanned the entire target object using the camera on an smartphone (Apple, iPhone 15) and a 3D scanning application (Epic Games, RealityScan - 3D Scanning App).
Suppose the measured 3D coordinates are represented in the world coordinate system.
We then selected 20 landmark points at distinctive geometric features such as corners and edges on the surface.
Next, we projected Gray-code patterns from each projector and used three ceiling-mounted cameras to capture the patterns, which were used to acquire the projector coordinate at each landmark.
From these correspondences, we computed the transformation matrix between each projector image coordinate system and the world coordinate system.
Accurate geometric registration of each projector onto the non-planar target was achieved by rendering the target object, with the desired texture mapped onto it, using a 3D computer graphics engine (OpenGL).
The geometric parameters of the virtual camera used for rendering were determined by the transformation matrix of each projector.

\begin{figure}[!tb]
  \centering
  \includegraphics[width=\linewidth]{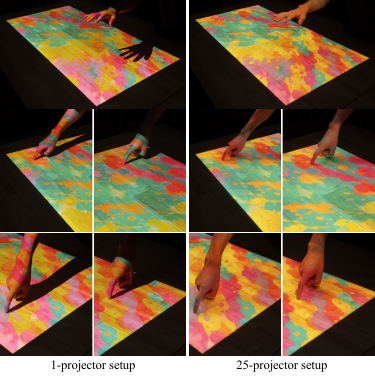}
  \caption{Comparison of shadow removal performance between single-projector (left) and 25-projector (right) setups on the tabletop surface. (Top) The hand is positioned approximately 0.2 m above the surface. (Bottom) The hand touches four different locations on the surface.}
  \label{fig:planar_result}
\end{figure}

\subsection{Evaluation of Shadow Removal Performance}\label{subsec:exp}

We evaluated the shadow removal performance for both planar and non-planar surfaces.
For comparison, we also built a single-projector setup (BenQ, TK850).
Because the 25-projector setup produced brighter projections, we applied a uniform scaling to the input images for all 25 projectors so that their combined brightness would match that of the single-projector setup.

\subsubsection{Planar Surface}\label{subsec:exp-planar}

In the experiment using a planar surface, images were projected onto a tabletop (size 1200 $\times$ 600 mm) placed approximately 1.4 m below the projector array.
Figure~\ref{fig:planar_result} (Top) shows the projection results when a user's hand was placed near the target surface (approximately 0.2 m away).
In the single-projector case, a clear shadow region appeared, whereas in the 25-projector case, no distinct shadow region was observed.
As shown in the supplementary video, where the hand is waved above the surface, there is no noticeable latency in the shadow removal process.
Figure~\ref{fig:planar_result} (Bottom) further shows the results when the user touched the surface with a fingertip.
While a noticeable shadow region appeared in the single-projector case, the 25-projector setup exhibited only slight luminance reduction, with no clearly defined shadows.
The figure also presents results from touching various locations across the table, demonstrating that effective shadow removal was achieved across a tabletop-sized workspace.

\begin{figure}[!tb]
  \centering
  \includegraphics[width=\linewidth]{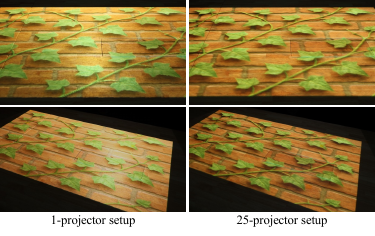}
  \caption{Comparison of specular reflections between single-projector (left) and 25-projector (right) setups on the tabletop surface, from two different viewing directions.}
  \label{fig:specular_reflection}
\end{figure}

An interesting phenomenon beyond shadow removal was also observed.
When using a single projector, strong specular reflections distorted the appearance of the projected content when viewed from certain angles.
In contrast, with the 25-projector setup, specular reflections were significantly suppressed (Fig.~\ref{fig:specular_reflection}).
Although the tabletop material was relatively matte, conventional PM often suffers from severe appearance degradation due to specular highlights caused by strong directional illumination.
In our 25-projector system, the intensity per projector was relatively low, which reduced specular reflections and provided a consistent visual appearance regardless of viewing angle.
Since projection targets are not always ideal diffuse surfaces, this characteristic of our synthetic aperture approach represents an important advantage.

\begin{figure*}[!bt]
  \centering
  \includegraphics[width=\linewidth]{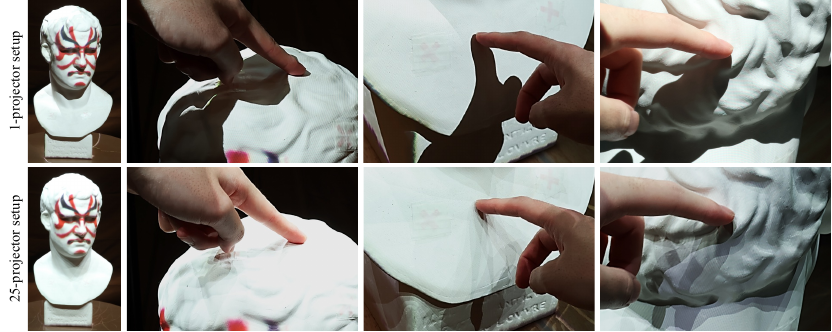}
  \caption{Comparison of shadow removal performance between single-projector (top row) and 25-projector (bottom row) setups on the plaster bust surface. (1st column) Projections without the occluder. (2nd--4th columns) The finger touches different parts of the surface: the head (2nd column), the bust (3rd column), and the lower back of the head (4th column).}
  \label{fig:3d_result}
\end{figure*}

\subsubsection{Non-Planar Surface}

As a non-planar target, we used a plaster bust (size $300\times250\times600$ mm, Fig.~\ref{fig:teaser} (Left)).
Figure~\ref{fig:teaser} (Center and Right) shows the projection results when a user's hand is placed near the surface.
While the single-projector setup caused the content to be occluded and disappear, the 25-projector setup maintained the projected content.
Figure \ref{fig:3d_result} shows the results when touching the target with a fingertip.
Shadow suppression was again more effective with 25 projectors.
However, for certain contact points, shadow regions larger than those observed on the planar surface were present, and their size varied depending on the location of contact.
We attribute this to the fact that the number of projectors illuminating each point on the target surface varies.

\begin{figure}[tb]
  \centering
  \includegraphics[width=\linewidth]{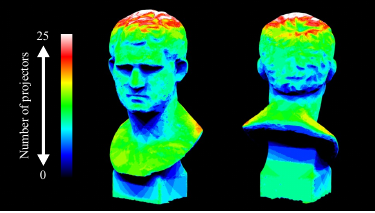}
  \caption{Pseudo-color visualization showing how many projectors illuminate different regions of the plaster bust surface.}
  \label{fig:3d_heatmap}
\end{figure}

Using the 3D shape data of the target surface and the projector positions estimated through the calibration process described in Section~\ref{subsec:system}, we computed the number of projectors contributing to each surface point, as visualized in Figure~\ref{fig:3d_heatmap}.
Comparison of the shadow area and the visualization confirmed that the shadow region became larger when fewer projectors contributed to a surface point.
Therefore, projectors must be arranged widely and densely to ensure that each point on the target surface is illuminated by a sufficiently large number of projectors.
However, determining the optimal number and density of projectors is a non-trivial task, as there is no established metric for evaluating sufficiency.
We address this issue through the user study described in Section~\ref{sec:user-study}.

\subsection{Evaluation of Blur Compensation Method}

We evaluated the proposed blur compensation method described in Section~\ref{sec:method}.
Specifically, we compared the computation time and compensation performance between the conventional single-projector blur compensation method from~\cite{10.1145/1141911.1141974}, naively extended to our multi-projector setting (Equation~\ref{eq:inverse_naive}), and our proposed method (Equation~\ref{eq:inverse_proposed}).

\begin{figure}[!tb]
  \centering
  \includegraphics[width=\linewidth]{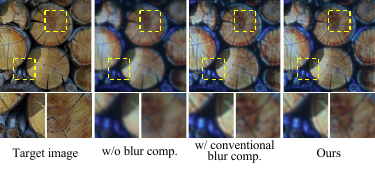}
  \caption{Blur compensation results. From left to right: the target image, the projected result without blur compensation, the result using conventional blur compensation, and the result using our proposed method.}
  \label{fig:blur_compensation}
\end{figure}

\begin{figure}[!tb]
  \centering
  \includegraphics[width=\linewidth]{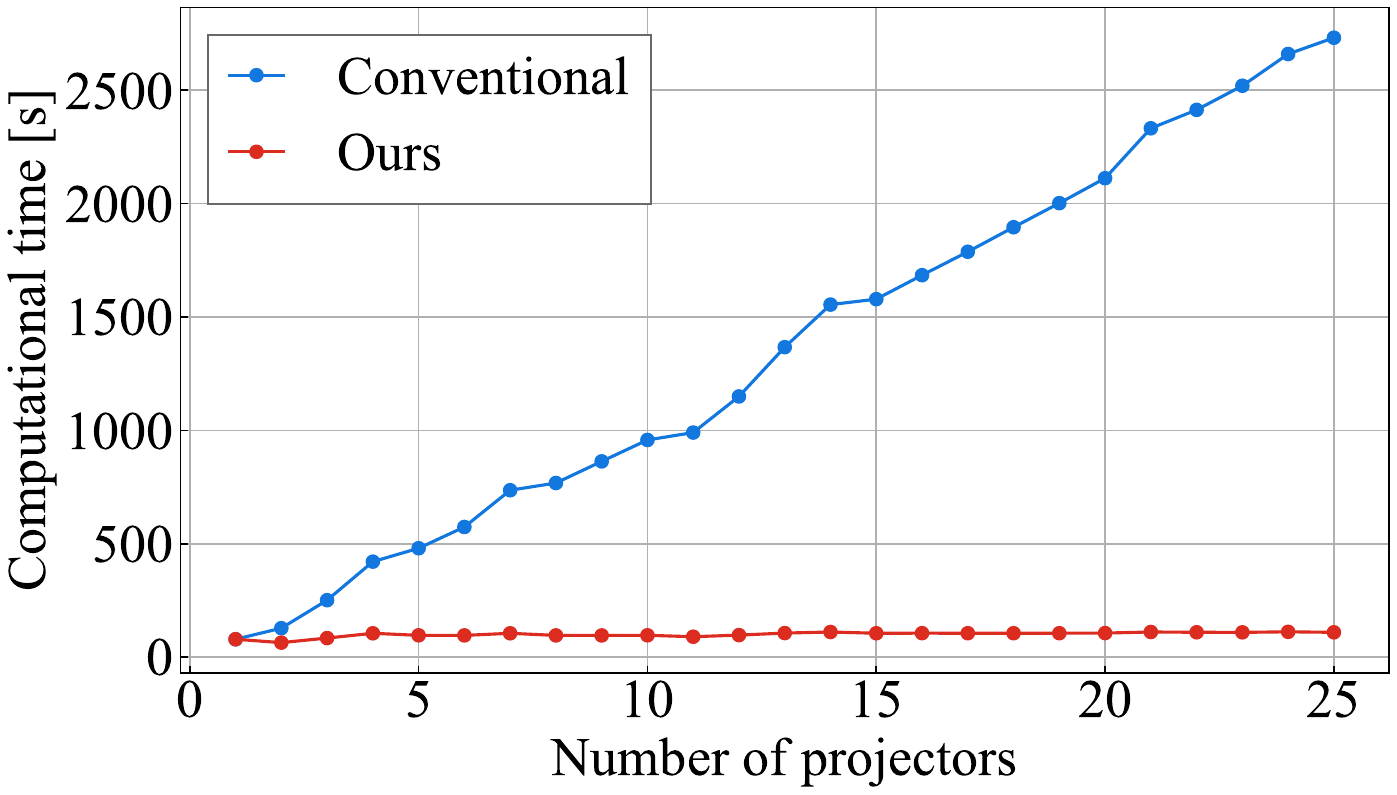}
  \caption{Computation time for generating the compensation image.}
  \label{fig:blur_compare}
\end{figure}

We used a planar surface, the same as in Section~\ref{subsec:exp-planar}, and set the distance between the projectors and the surface to 1.4 m.
Figure~\ref{fig:blur_compensation} shows the target image, the projected result without blur compensation, the result using conventional blur compensation, and the result using our proposed method.
The projection result without blur compensation appear significantly blurred, whereas the results with the two blur compensation methods show an improvement in image sharpness.
In practice, small residual subpixel misalignments can remain after calibration, which may leave slight edge blur even after compensation.
We computed PSNR and SSIM to quantitatively evaluate the error between the target and the projection results and summarized the results in Table~\ref{table:blur_evaluation}.
We also compared the computation times, as shown in Table~\ref{table:blur_evaluation} and Figure~\ref{fig:blur_compare}.
These results demonstrate that our method suppresses image degradation as effectively as the conventional approach, while reducing computation time by a factor of 25, corresponding to the number of projectors.

\begin{table}[!tb]
    \centering
    \caption{Similarity between the target appearance and projected results, and processing time.}
    \label{table:blur_evaluation}
    \begin{tabular}{cccc} 
    \hline
   &PSNR($\uparrow$)&SSIM($\uparrow$)&times [s]\\ \hline
   w/o blur comp.& 19.54 & 0.44 \\
   w/ conventional blur comp.& 20.00 & 0.49 & 2732\\
   Ours& \textbf{20.05} & \textbf{0.50} & 109\\ \hline
   \end{tabular}
\end{table}

\section{User Study}\label{sec:user-study}

\begin{figure}[!tb]
  \centering
  \includegraphics[width=\linewidth]{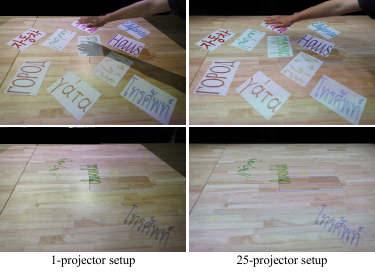}
  \caption{Stimuli used in the first user study. Ten sheets of paper were placed on the tabletop surface: seven had printed text in various languages, and the remaining three were blank with projected text patterns instead.}
  \label{fig:us1_setup}
\end{figure}

To design a synthetic aperture projection system with sufficient shadow-removal performance, it is essential to determine both the number and spatial arrangement of projectors.
While increasing the number of projectors directly improves coverage, it also incurs higher economic and computational costs, making it desirable to minimize the number used.
Regarding spatial arrangement, distributing projectors over a wide area helps accommodate various occlusion scenarios.
However, in the presence of occluders, such wide distribution can also lead to more extensive light drop regions.
These factors present a trade-off, making it non-trivial to determine the optimal configuration.
Moreover, there is currently no established metric for evaluating shadow-removal performance in PM, which further complicates the determination of the optimal number and placement of projectors.

During informal user studies conducted using the system described in Section~\ref{subsec:system}, we observed that when a large number of projectors were densely arranged, significantly reducing light drop, participants no longer perceived the changes in appearance as being caused by overlaid imagery.
Instead, the changes were interpreted as natural variations in the surface's reflectance properties, indicating a suppression of the SoP.
Suppressing the SoP and evoking the perception of altered material reflectance is essential in various application domains such as industrial design~\cite{CASCINI2020103308,10937893,8797923}, restoration of deteriorated cultural heritage~\cite{10.1145/1409060.1409102}, and cosmetics~\cite{10316453}.
Based on this insight, we adopt the SoP as a metric for evaluating shadow-removal performance.

This section reports two user studies.
The first investigates whether the proposed system can suppress the SoP.
The second explores the relationship between SoP and the number and spatial arrangement of projectors, aiming to establish design guidelines for synthetic aperture projection systems.
Note that these studies were approved by the institutional review board (IRB) of the author's institution (Approval ID: R6-19).

\subsection{Demonstrating the Suppression of the Sense of Projection}\label{subsec:study1}

\begin{figure}[!tb]
  \centering
  \includegraphics[width=\linewidth]{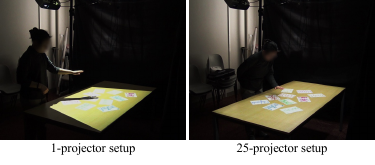}
  \caption{Participants performing the task under the 25-projector condition in the first user study. The participant’s face has been intentionally blurred to protect their identity.}
  \label{fig:us1_participants}
\end{figure}

To demonstrate that SoP is suppressed and that the appearance is perceived as a change in material reflectance properties, we conducted an experiment using the 25-projector setup described in Section~\ref{subsec:system}.
We used the full 5$\times$5 projector configuration as a representative setting of our prototype system.
In Section~\ref{subsec:study2}, we further investigate how SoP depends on projector number and arrangement, including reduced configurations.
Specifically, we placed ten A5-sized sheets of paper on a table without overlapping.
Among them, seven sheets had printed texts in various languages, and the remaining three sheets were blank and displayed projected text patterns instead (Fig.~\ref{fig:us1_setup}).
To reduce the sharpness mismatch between printed and projected text, we additionally blurred the printed text stimuli so that their perceived sharpness was comparable to that of the projected text.
Participants were instructed to identify the three sheets with projected text patterns.
No strict time limit was imposed; they were asked to respond as quickly as possible while maintaining accuracy.
They were allowed to move freely around the table and block the projected light with their bodies, as long as they did not touch the sheets (Fig.~\ref{fig:us1_participants}).
No formal training was provided and they received only brief task instructions before starting.
There were two experimental conditions: one using the 25-projector setup and one using a single projector.
We predict that, if SoP is strongly suppressed in a condition, the projected characters become perceptually indistinguishable from the printed ones.
This increases the difficulty of the identification task and resulting in longer task completion times and lower accuracy.
As in the experiments in Section~\ref{subsec:exp}, the pixel values of the projected images were linearly scaled to match the illuminance on the target surface across both conditions.
A uniform gray pattern (33.0 $cd/m^2$) was set as the background of the projected images.
To ensure that the single-projector condition achieved comparable brightness, we used a high-luminance projector (BenQ, TK850).

\begin{figure}[!tb]
  \centering
  \includegraphics[width=\linewidth]{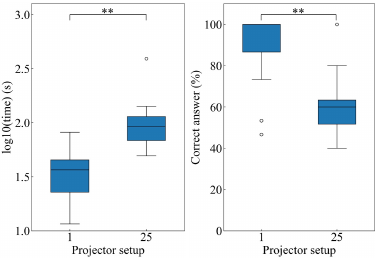}
  \caption{Average task completion time (left) and accuracy (right) for each condition in the first user study.}
  \label{fig:us1_result}
\end{figure}

Seventeen participants (age 19 to 28; 8 male, 9 female) took part in the experiment.
Each participant performed the task 10 times under both conditions.
Figure~\ref{fig:us1_result} shows the average task completion time and accuracy for each condition.
We quantified accuracy as the proportion of the three projections detected correctly in each trial. Concretely, the accuracy score was 0\%, 33.3\%, 66.7\%, or 100\%, depending on whether 0, 1, 2, or all 3 projections were detected correctly.
For the average time to complete the task on a log scale, because normality assumption was not violated (Shapiro-Wilk's normality test, \textit{p} $>$ .05), we conducted a paired \textit{t}-test. The result showed that the time was significantly higher in the 25-projector condition than in the single-projector condition (\textit{t} (16) = 4.87, \textit{p} $<$ .01, Cohen's \textit{d} = 1.85).
For accuracy, we conducted an Wilcoxon signed-rank test. The result showed that accuracy was significantly higher in the single-projector condition than in the 25-projector condition (\textit{p} $<$ .01, Cohen's \textit{r} = 0.50).
The 25-projector condition resulted in task times $10^{1.3}$ times longer and accuracy 0.7 times lower than the single-projector condition.
These results demonstrate that the proposed shadowless PM system is capable of suppressing SoP, allowing the projected content to be perceived as variations in the surface's reflectance properties.

\subsection{Effect of the Number and Spatial Arrangement of Projectors on the Sense of Projection}\label{subsec:study2}

\begin{figure}[!tb]
  \centering
  \includegraphics[width=\linewidth]{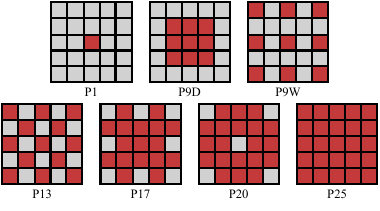}
  \caption{Spatial arrangements of projectors in the tested configurations. Red cells indicate the projectors used in each configuration.}
  \label{fig:us2_pro_condition}
\end{figure}

We conducted another user study to investigate how the number and spatial arrangement of projectors affect SoP.
Participants were shown paired projected results under different configurations and asked two questions for each pair: (1) which image appeared more as a pattern intrinsic to the surface (i.e., attributable to material reflectance), and (2) which image appeared more as a projected pattern.
Lower SoP corresponds to stronger perception of the former and weaker perception of the latter.
We prepared seven configurations, which were labeled P1, P9D, P9W, P13, P17, P20, and P25 (Fig.~\ref{fig:us2_pro_condition}).
P1 served as the baseline using a single projector.
A high-luminance projector (the same as in Section~\ref{subsec:study1}) was used to ensure brightness comparable to multi-projector setups.
The remaining configurations used subsets of a 5$\times$5 array of 25 projectors described in Section~\ref{subsec:system}.
P9D and P9W used 9 projectors each.
P9D employed the projectors placed densely at the center, while P9W used a wider distribution (1 center and 8 outer projectors).
P13, P17, P20, and P25 used increasing numbers of outer projectors (13, 17, 20, and 25), gradually increasing spatial density.

In this study, we formulated the following hypotheses:
\begin{description}
    \item[H1] Suppressing shadows reduces sense of projection (SoP).
    \item[H2] A wider arrangement improves shadow removal and thus suppresses SoP.
    \item[H3] Increasing the number of projectors reduces the light drop when occluded, thereby suppressing SoP.
\end{description}

\begin{figure}[!tb]
  \centering
  \includegraphics[width=\linewidth]{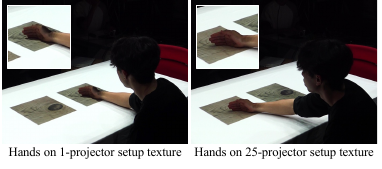}
  \caption{Participants observing two projected images under different projector configurations, while placing their hands over each one.}
  \label{fig:us2_participant}
\end{figure}

\begin{figure}[!tb]
  \centering
  \includegraphics[width=\linewidth]{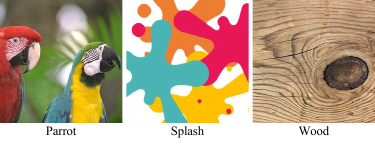}
  \caption{Projection images used in the second user study.}
  \label{fig:us2_img_condition}
\end{figure}

The projection target was a table covered with a plain white cloth, placed horizontally at a distance of 1.4 m from the projector array.
As shown in Figure~\ref{fig:us2_participant}, two projected images under different projector configurations were presented side by side (left and right) to each participant for comparison.
To help participants observe how shadows were cast, they were instructed to place their left and right hands sequentially over each of the two projected images (Fig.~\ref{fig:us2_participant}).
After this, they were asked to answer two questions using a 5-point scale: (1) which side appeared more as a pattern intrinsic to the surface (-2: clearly left, -1: slightly left, 0: equal, 1: slightly right, 2: clearly right), and (2) which side appeared more as a projected pattern, using the same 5-point scale.
We included both questions to verify that the two judgments form a complementary pair and to check that the results are robust to question wording.
Three types of projection images were used for this task (parrot, splash, and wood; see Fig.~\ref{fig:us2_img_condition}).
Non-image areas were filled with a uniform gray projection (22.4 $cd/m^2$).

\begin{figure*}[!tb]
\centering
\subfloat[]{\includegraphics[width=0.5\linewidth]{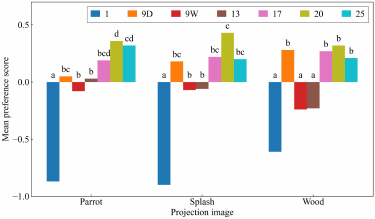}%
\label{fig:us2_result_q1}}
\hfil
\subfloat[]{\includegraphics[width=0.5\linewidth]{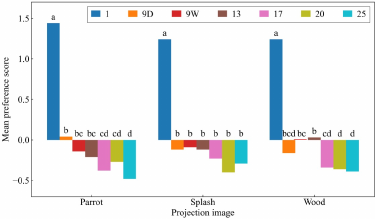}%
\label{fig:us2_result_q2}}
\caption{Mean preference scores and statistical test results obtained using Scheff\'{e}'s method (Nakaya’s modification) for pairwise comparisons. Bars labeled with different letters are significantly different at \textit{p} $<$ .01.  (a) Question 1: Which side appeared more as a pattern intrinsic to the surface? (b) Question 2: Which side appeared more as a projected pattern?}
\label{fig:us2_result}
\end{figure*}

Seventeen participants (age 19 to 26; 8 male, 9 female) took part in the study.
Note that the participants in this user study were different from those in the previous study.
Prior to the experiment, participants were briefed on the perceptual difference between patterns intrinsic to the surface (i.e., due to surface reflectance) and projected patterns.
For each of the three projection images, each participant completed one pairwise comparison for every combination of the seven projector configurations.
Figure~\ref{fig:us2_result} presents the mean preference scores and statistical test results obtained using Scheff\'{e}'s method (Nakaya's modification) for pairwise comparisons~\cite{nakaya1970scheffe}. 
Note that Scheff\'{e}'s method yields a relative psychological scale of SoP across the tested configurations, rather than an absolute scale with a physical reference.
Accordingly, the scale values should be interpreted as relative magnitudes of SoP within this comparison set, not as an absolute distance to the printed one.
Bars with different letters are significantly different at \textit{p} $<$ .01.
Details of the statistical analysis, including the yardstick $Y$ used to compute confidence intervals for pairwise differences, can be found in the supplementary materials.

The results indicate that all six multi-projector configurations were significantly more likely to be perceived as patterns intrinsic to the surface and less likely as projected patterns, compared to the single-projector baseline (P1).
This supports hypothesis H1, confirming that suppressing shadows using multiple projectors reduces SoP.
When comparing P9D and P9W, which both use nine projectors but differ in distribution density, P9D (denser and more localized) resulted in a stronger perception of surface reflectance patterns.
This difference was statistically significant for the wood image.
These findings suggest that increasing projector density is more effective than expanding the coverage area in reducing SoP, thereby contradicting hypothesis H2.
Next, no significant differences in SoP ratings were observed among the P17, P20, and P25 conditions.
This suggests that beyond a certain number of projectors within a fixed coverage area, additional projectors do not further reduce the SoP, failing to support hypothesis H3.
Interestingly, although not statistically significant, the P20 condition yielded better SoP suppression than P25.
In P20, the central projector was excluded.
Since the central projector, being directly facing the target surface, typically contributes the highest brightness but is also most likely to be occluded, its absence in P20 likely reduced light drop artifacts, leading to improved perception.

\section{Discussion}

In this study, we demonstrated delay-free, shadowless PM for tabletop-sized workspaces using synthetic-aperture projection with multiple projectors.
With our prototype system consisting of 25 projectors arranged in a 2D array, we confirmed high shadow-removal performance on planar surfaces facing the projector array.
We also confirmed that shadow removal is achievable on non-planar surfaces.
Furthermore, we found that, when projecting onto non-planar surfaces, the number of contributing projectors significantly affects shadow-removal performance.
To address the blurring that occurs due to overlapping projections from multiple units, we proposed an efficient compensation method and confirmed its effectiveness through an experiment.

Furthermore, we conducted user studies to establish guidelines for determining the optimal number and spatial arrangement of projectors.
We introduced the SoP as a metric to evaluate shadow-removal quality and demonstrated that our 25-projector system significantly suppresses SoP, allowing users to perceive changes as alterations in surface reflectance rather than projections.
Through comparative evaluations of various configurations, we found that to suppress SoP and enable shadow removal that is perceived as changes in surface material reflectance properties, projectors should be arranged such that each point on the target surface is illuminated by multiple projectors.
Based on the results, we derived the following design guidelines: (1) approximately 17 projectors are sufficient, (2) the distance between projectors should not be too large to avoid sparse coverage, and (3) projectors should not be placed in positions that are highly likely to be occluded.

Future work includes computing optimal projector placements based on the geometry and orientation of the target object to minimize SoP.
For example, when targeting non-planar surfaces, these findings suggest that projectors should be installed not only on the ceiling but also on surrounding walls.
Conversely, when the projector layout is fixed, determining the optimal position and orientation of the target object to achieve minimal projection artifacts will also be important.

One limitation of our current approach is the need for a large number of projectors.
While we used 25 projectors in the experiments, even more may be necessary for projecting onto complex 3D surfaces.
Our prototype system also requires special image processing equipment such as video signal splitters, and the number of projectors that can be connected to a single PC is limited, making the system unscalable.
To address this, we envision a distributed control framework~\cite{7164338} in which each projector is connected to a low-power edge computer for independent video rendering.
This would allow for a more scalable projection system.
Additionally, increasing the number of projectors lengthens the time required for geometric calibration, as each projector must sequentially project gray code patterns.
To mitigate this, we consider it a promising future direction to apply recent techniques that decompose projected light fields using image sensors as calibration probes~\cite{9523844}.
This approach could enable simultaneous separation and measurement of gray code patterns projected from multiple projectors, drastically reducing calibration time.

\section{Conclusion}

This paper presented a novel PM system that achieves delay-free and shadowless rendering for tabletop workspaces using a synthetic-aperture projection approach. 
By densely arranging 25 projectors in a 2D array, we successfully demonstrated high-performance shadow removal even in dynamic interaction scenarios.
To address the blurring that occurs from overlapping projections, we proposed an efficient compensation technique that maintains high image fidelity while remaining computationally scalable.
Furthermore, we introduced the SoP as a perceptual metric to evaluate shadow-removal effectiveness, and through user studies, validated that multi-projector configurations can significantly reduce SoP.
These findings allowed us to derive practical guidelines for the number and placement of projectors, revealing that under our tabletop experimental setup, approximately 17 units in a moderately dense configuration are sufficient to achieve perceptual realism.
We also identified conditions under which SoP suppression is most effective, such as avoiding occlusion-prone positions and maintaining appropriate spacing between projectors.
While our prototype demonstrates promising results, it currently requires a large number of projectors and specialized hardware.
We discussed future technical improvements, including distributed rendering systems and faster geometric calibration methods, to enhance scalability and usability.
As a future direction, we plan to deploy our system in specific application fields---such as industrial design support---and conduct real-world experiments to evaluate its practical effectiveness.








\acknowledgments{%
	The authors wish to thank Sora Kawashima, Yuto Tateiwa, and Hiroki Kusuyama.
  This work was supported by JSPS KAKENHI, Japan Grant Number JP25K03155, JP25K22820, JST ASPIRE, Japan Grant Number JPMJAP2404, and JST BOOST, Japan Grant Number JPMJBS2402.
}

\bibliographystyle{abbrv-doi-hyperref}

\bibliography{template}

\appendix 

\section{About Appendices}
Refer to \cref{sec:appendices_inst} for instructions regarding appendices.

\section{Troubleshooting}
\label{appendix:troubleshooting}

\subsection{ifpdf error}

If you receive compilation errors along the lines of \texttt{Package ifpdf Error: Name clash, \textbackslash ifpdf is already defined} then please add a new line \verb|\let\ifpdf\relax| right after the \verb|\documentclass[journal]{vgtc}| call.
Note that your error is due to packages you use that define \verb|\ifpdf| which is obsolete (the result is that \verb|\ifpdf| is defined twice); these packages should be changed to use \verb|ifpdf| package instead.

\subsection{\texttt{pdfendlink} error}

Occasionally (for some \LaTeX\ distributions) this hyper-linked bib\TeX\ style may lead to \textbf{compilation errors} (\texttt{pdfendlink ended up in different nesting level ...}) if a reference entry is broken across two pages (due to a bug in \verb|hyperref|).
In this case, make sure you have the latest version of the \verb|hyperref| package (i.e.\ update your \LaTeX\ installation/packages) or, alternatively, revert back to \verb|\bibliographystyle{abbrv-doi}| (at the expense of removing hyperlinks from the bibliography) and try \verb|\bibliographystyle{abbrv-doi-hyperref}| again after some more editing.

\end{document}